\documentstyle[12pt]{article}
\def\ba{\begin{eqnarray}}
\def\ea{\end{eqnarray}}

\def\lb{\label}
\def\be{\begin{equation}}
\def\ee{\end{equation}}

\begin{document}

\title{A construction of $G_2$ holonomy spaces with torus symmetry}
\author{O.P. Santillan \thanks{Telephone:(7-09621) 65-400, Fax: (7-09621) 65-084, e-mail: osvaldo@thsun1.jinr.ru} \\
Bogoliubov Laboratory of Theoretical Physics, JINR,\\
141 980 Dubna, Moscow Region, Russia}
\date {}
\maketitle

\begin{abstract}
In the present work the Calderbank-Pedersen description of four dimensional
manifolds with self-dual Weyl tensor is used to obtain examples
of quaternionic-kahler metrics with two commuting isometries.
The eigenfunctions of the hyperbolic laplacian are found
by use of Backglund transformations acting over
solutions of the Ward monopole equation.
The Bryant-Salamon construction of $G_2$ holonomy metrics arising as $R^3$
bundles over quaternionic-kahler base spaces is applied to this examples
to find internal spaces of the M-theory that leads to an $N=1$ supersymmetry
in four dimensions. Type IIA solutions will be obtained too by reduction along
one of the isometries. The torus symmetry of the base spaces
is extended to the total ones.
\end{abstract}

\section{Introduction and the main result.}

    The classification of the possible holonomy groups of
Riemannian or Pseudo-Riemannian manifolds is an old mathematical
problem. In \cite{Berger} Berger presented a list of the possible
restricted holonomy groups of N-dimensional Riemannian manifolds, but
after the completion of this work it remained to prove the existence
of metrics with exceptional holonomies $G_2$ and $Spin(7)$ for the seven and
eight dimensional cases respectively. This was achieved successfully by
Bryant in \cite{Bryant}. In general, if a given Riemannian metric
with dimension $N$ admits at least one covariantly constant spinor
satisfying $D_i\eta=0$ the holonomy group will be $SU(\frac{n}{2})$
, $Sp(\frac{n}{4})$, $G_2$ or $Spin(7)$, the last two cases corresponding to
seven and eight dimensions. For $G_2$ holonomy manifolds there is exactly one.
The fact that this spinor exist is apparent
from the decomposition $8=1 \oplus 7$ of the spinor representation of the tangent space
$SO(7)$. Equivalently, in such manifolds one can choose an
orthogonal frame $e^i$ in which the three octonionic form
$$
\Phi=e^1 \wedge e^2 \wedge e^7 + e^1 \wedge e^3 \wedge e^6 + e^1
\wedge e^4 \wedge e^5 + e^2 \wedge e^3 \wedge e^5 + e^4 \wedge e^2
\wedge e^6
$$
$$
+\; e^3 \wedge e^4 \wedge e^7 + e^5 \wedge e^6 \wedge e^7
$$
and its dual $\ast\Phi$ are closed \cite{Gibb2}.

      After the Bryant work explicit compact and non-compact $G_2$ holonomy metrics
were constructed in \cite{Lolo} and \cite{Salamon}, and complete ones in \cite{Gomis} and \cite{Branbauer}.
Recently, new examples have been found in \cite{Joyce3},
\cite{Konishi}, \cite{Gibbi}, \cite{Gibbano}, \cite{Chong}, \cite{Lu}
and \cite{Gibbe}. All of them have
vanishing Ricci tensor, i.e, they are Ricci-flat, and this implies that they
are vacuum solutions of the Einstein equation.
This suggest that the construction of $G_2$ holonomy manifolds is related with the
construction of seven dimensional
self-dual manifolds. The self-duality condition for the spin connection
in seven dimensions
$$
w_{ab}=\pm\frac{c_{abcd} }{2}w_{cd}
$$
implies Ricci-flatness and $G_2$ restricted holonomy \cite{Acharya} (analogous
considerations hold in eight dimensions for
$Spin(7)$). Given an anzatz for a metric,
this condition gives a system of equations to be satisfied
in order to have a $G_2$ manifold. Although this equations are non-linear
they have been solved in cases with suitable symmetries in which the system
takes a more simple form \cite{Gibb2}.

   The main feature that relates this subject to physics is the
presence of one nonzero parallel spinor field
$\eta$, which plays a central role in supersymmetry, string theory and
M-theory. Compactification of the M-theory (or his low energy limit, the eleven dimensional
supergravity) on $G_2$ holonomy manifolds leads to an effective four dimensional theory with
one supersymmetry, corresponding to such spinor field \cite{Papadopolus}.

   There are in the literature examples of "weak
$G_2$ holonomy" \cite{Duff}, which are again backgrounds of the M-theory
that give rise to N=1 supersymmetry in $D=4$.
In this case, there is an spinor field $\eta$ which is not
covariantly constant but satisfies $D_i\eta\sim\lambda\gamma_i \eta$.
The Ricci flatness condition is replaced by $R_{ij}\sim \lambda g_{ij}$.
In the limit $\lambda\rightarrow 0$ one obtain $G_2$ as restricted holonomy group.
Hitchin has shown that under certain conditions is possible to construct this kind of
manifolds starting with an $Spin(7)$ holonomy one \cite{Hitchin}.

   Compactification of the M-theory based on $G_2$ smooth manifolds cannot give rise to chiral matter.
In the smooth case the harmonic Kaluza-Klein decomposition of the eleven-dimensional
supergravity is the N=1 four dimensional supergravity coupled Abelian vector multiplets
plus chiral multiplets. But the chiral matter fields can emerge only if the manifold
develops a singularity, as pointed out by Witten and Acharya in \cite{Acharya2} and \cite{Witten2}.
It turns out that to obtain a realistic model one should investigate the dynamics
of the M-theory over orbifolds. A modern description of this dynamics over manifolds
that are developing a conical singularities can be found in \cite{Witten}. Generalizations
of the work of Acharya and Witten on singular $G_2$ spaces were investigated recently in
\cite{Berglund}, and over spaces with torus symmetry and
only orbifold singularities in \cite{Angelova} and \cite{Angelito}.

     Certain $G_2$ metrics with two abelian isometries have called the attention
recently \cite{Mahapatrah}, because an $U(1)$ isometry
allows an type IIA superstring interpretation upon dimensional reduction
to ten dimensions. In \cite{Angelova} Anguelova and Lazaroiu have extracted the
IIA reduction of M-theory on certain toric backgrounds, and its type IIB duals.
Such IIA solutions corresponds to systems of weakly and strongly coupled D6-branes,
while the duals describes systems of localized and delocalized 5-branes.
The $G_2$ spaces presented in those works have been found applying the Bryant-Salamon
construction with a four dimensional quaternionic base manifold with torus
symmetry \cite{Salamon}; the $U(1)\times U(1)$ isometry is extended to the total
space.

     The four dimensional toric quaternionic kahler spaces has been completely described by Calderbank
and Pedersen \cite{Pedersen} in terms of eigenfunctions of the hyperbolic laplacian
with eigenvalue $3/4$, namely the solutions of the equation
$$
F_{\rho\rho}+F_{\eta\eta}=\frac{3F}{4\rho^2}.
$$
The toric $G_2$ cones utilized in \cite{Angelova} were constructed with the so called
$m$-pole eigenfunctions which gives rise to spaces that are complete (thus compact)
and admits only orbifold singularities. As such, they seem to be the best candidates
with $U(1)\times U(1)$ isometry for which the physical analysis of Acharya
and Witten can be applied.

     One of the purposes of this paper is to describe how to construct non trivial eigenfunctions
of the hyperbolic laplacian (and, in consequence, examples of toric quaternionic spaces) starting with
solutions of the monopole equation
$$
V_{\eta\eta}+\rho^{-1}(\rho V_{\rho})_{\rho}=0
$$
described by Ward and Woodhouse in \cite{Ward}. In fact, in \cite{Pedersen} it has been pointed out that
the space of solutions of the Ward monopole equation and the Calderbank Pedersen one are
related by a Backglund transformation, allowing to construct solutions one from another. The Ward monopole
equation has been investigated in the context of (2+1) gravity and it is known that its solutions admits
an integral representation in terms of an arbitrary function of one variable. For this reason it is possible
to find non trivial eigenfunctions of the hyperbolic laplacian selecting a function in the integral
representations of the monopoles and performing a Backglund mapping.

    The Ward monopoles are also relevant for hyperkahler geometry \cite{Calderbank2}.
This is because it is possible to construct hyperkahler spaces with torus symmetry starting with
a given monopole, in analoguous manner as in the Calderbank Pedersen picture. Moreover,
in the hyperkahler limit the Bryant Salamon construction gives $G_2$ holonomy spaces which are
globally the cartesian product of the hyperkahler one with $R^3$.
For this reason one can construct such spaces starting with a monopole,
without using a Backglund mapping. The reason for which
in the hyperkahler limit it is obtained
a trivial product with $R^3$ could be easy to visualize: being $R^3$ flat the seven dimensional self duality
condition becames the four dimensional one for the base space.

   The organization of this paper is as follows: in section 2 it is presented
the group $G_2$ as the group of automorphisms of the octonions. In
section 3 there are reviewed some basic features about self-dual seven dimensional
manifolds. In section 4 it is shown that such manifolds have restricted holonomy group $G_2$.
Section 5 contains a review of the Bryant Salamon construction
of $G_2$ metrics with quaternionic kahler base spaces.
In section 6 the Ward and Calderbank-Pedersen descriptions are reviewed and
examples of quaternionic kahler and hyperkahler metrics are constructed.
The $m$-pole solutions are also discussed in some detail.
The vacuum configurations of the eleven dimensional supergravity
related to this metrics are founded in section 7. Dimensional reduction
of such configurations alone one of the isometries is performed to
give rise to ten dimensional type IIA backgrounds.

  To conclude, it should be mentioned that the study of explicit
metrics with exceptional holonomy has also importance in the context of dualities
of string theory and M-theory (\cite{Vafa}-\cite{Kachru}). The range of
applications of this topic is very wide; some of them
can be found in \cite{Partuche}-\cite{Acho}.

\section{The exceptional group $G_2$ and the octonions.}

  The group $G_2\subset SO(7)$ is one of the exceptional simple Lie groups. It
is compact, connected, simply-connected and of dimension $14$. It
has been proved \cite{Schafer}, that $G_2$ is the group of automorphisms
of the octonions $O$, up to an isomorphism. The octonions (or Cayley numbers) constitutes the
only non associative division algebra (the associative ones are only $R$, $C$ and $H$)
and an arbitrary element $x \in O$ can be written as a linear
combination of the form $\overrightarrow{x}=x^0 + x^i e_i$, where the set $e_i$
constitute a basis of $7$ unit octonions with the following multiplication rule:
\be\lb{mult}
e_ie_j=c_{ijk}e_k;\;\;\; e_i.1=1.e_i=e_i
\ee

The $x^i$'s takes real values. The subspace $P$ of $O$ generated by the
elements $x= x^i e_i$ is called the space of "pure octonions", and the total
space can be decomposed as $O=R \bigoplus P$. The constants $c_{ijk}$ that
define the multiplication (\ref{mult}) are totally antisymmetric and
\be\lb{const}
c_{123}=c_{246}=c_{435}=c_{367}=c_{651}=c_{572}=c_{714}=1,
\ee
up to an index permutation. The constants corresponding to another set of indices are
identically zero. From (\ref{mult}) and (\ref{const}) it is seen that
$(e^3e^7)e^5-e^3(e^7e^5)=-e^1$,
which shows the non-associativity of the octonion algebra. For this reason
the octonions cannot be represented as a matrices and do not satisfy
the Jacobi identities. In other words, the algebra of $O$ is not a Lie
algebra.

   It is possible to represent the components of an arbitrary octonion as
a $7$-dimensional vector $\overrightarrow{X}=(x_1,...,x_7)$. We define the octonion numbers
$g\overrightarrow{x}$ as those with components $g.\overrightarrow{X}$, where $g$ is an
arbitrary $7 \times 7$ matrix. The statement that the exceptional group $G_2$
is the group of automorphism of the octonions means if $x \in P$, $gx \in P$
if $g$ is any of the elements of $G_2$ in the fundamental irreducible representation;
and that if $x.y=z$ for given x, y and z belonging to $P$, $gx.gy=gz$.

  Over $O$ it is defined an internal product $(,):O\times O \rightarrow
R$ given by

\be\lb{intprod} (e_i,e_j)=\delta_{ij}\;, \ee

from where it is obtained that
 \be\lb{intprod2} (x,y)=x^iy^i\;. \ee

  Taking into account all the facts mentioned above it is possible
to construct a three-form over a seven dimensional space V, which is $G_2$
invariant. This form is fundamental in this work because its
closure has implications about the holonomy group of V. The construction
follows decomposing the octonion space as $O=R
\bigoplus P$ , and defining over $P$ the bilinear $B(x,y)$ and the
internal product $<,>:P\times P\rightarrow R$ by the identities

\be\lb{redprod} (x^0e_0 + x, y^0e_0 + y)=x^0y^0 + <x,y>\;, \ee
\be\lb{bilin} (x^0e_0 + x)(y^0e_0 + y)=(x^0y^0 - <x,y>)e_0 +(x^0y
+ y^0x + B(x,y))\;. \ee

Under an automorphism transformation $B(x,y)$ and $<,>$ satisfy

\be \lb{1equi} B(gx,gy)=gB(x,y)\;, \ee \be \lb{2equi}
B(x,y)=-B(y,x)\;, \ee \be \lb{3equi} <gx,gy>=<x,y>\;. \ee

From (\ref{1equi}) and (\ref{3equi}), it is seen that \be\lb{trili}
\Phi(x,y,z)=<B(x,y),z>=<B(gx,gy),gz>=\Phi(gx,gy,gz)\;. \ee

In other words, the trilinear $\Phi(x,y,z)=<B(x,y),z>$ is $G_2$
invariant. From the rule (\ref{mult}) and the definition (\ref{bilin}) it follows
that components of $\Phi$ are
$$
\Phi(e_a, e_b, e_c)=<B(e_a, e_b),e_c>=c_{abc}\;,
$$
and that
$$
\Phi(x,y,z)=c_{abc}x^ay^bz^c.
$$
 From the invariance of $\Phi$ under the action of $G_2$ it follows
that for a given real vector space V of dimension $7$, with $e_1,..,e_7$ a
basis for V, the three form
\be\lb{3form}
\Phi(x,y,z)=\frac{1}{3!}c_{abc}e^a \wedge e^b \wedge e^c
\ee
is $G_2$ invariant.

\section{Self-dual manifolds in $4$ and $7$ dimensions.}

 The self-duality condition is a familiar concept in quantum field theory
\cite{Belavin} and in general relativity. By definition the curvature tensor
$R_{ab}=dw_{ab}+w_{ac}\wedge w_{cb}$ of a Riemannian metric is self-dual if
\be\lb{4duality}
R_{ab}=\frac{1}{2}\epsilon_{abmn}R_{mn},
\ee
or, in components
$$
R_{abcd}=\frac{1}{2}\epsilon_{abmn}R_{mncd}.
$$
Non trivial solutions of this type has been found in the past
\cite{Eguchi}. They are called
"gravitational instantons" if their are Euclidean and with finite energy.
It has been shown that in four dimensions a torsion-free metric that
satisfies (\ref{4duality}) will be a vacuum solution of the
Einstein equations without cosmological constant \cite{Eguchi2}.

  In seven dimensions (\ref{4duality}) is generalized
 as \cite{Acharya}

 \be\lb{7duality} R_{ab}=\frac{1}{2}c_{abcd}R_{cd}\;. \ee

 where the totally antisymmetric $c_{abcd}$ are defined in terms
of the octonion structure constants (\ref{const}) through the relations:

\be\lb{const1} c_{abcd}=\frac{1}{3!}\epsilon^{abcdefg}c_{efg}\;.
\ee

This generalization has been related to the octonions
because the solutions of (\ref{7duality}) will be not only vacuum
solutions of the Einstein equation without cosmological constant,
but Ricci-flat, which is one of the main features of the $G_2$ manifolds.
This follows directly from (\ref{7duality}), the antisymmetry of (\ref{const1}) and
the Bianchi identity $R_{d[ebc]}=0$ as
\be\lb{flat}
R_{ab}=R_{acbc}=\frac{1}{2}c_{acde}R_{debc}=\frac{1}{2}c_{acde}R_{d[ebc]}=0\;.
\ee
Indeed, it will be shown in the next section that seven dimensional self-dual
manifolds have restricted holonomy $G_2$. For such case the Weyl tensor
$C_{abcd}$ defined by
\be\lb{weyl}
C_{abcd}=R_{abcd}+\frac{R}{(n-1)(n-2)}(g_{ac}g_{bd}-g_{ad}g_{bc})
-\frac{1}{(n-2)}(g_{ac}R_{bd}-g_{ad}R_{bc}-g_{bc}R_{ad}+g_{bd}R_{ac})
\ee
will be equal to the Riemann tensor. The first one is traceless,
in consequence Ricci-flat manifolds have traceless curvature tensor.

   Many authors presents the seven dimensional self-duality as a property
of the spin connection, namely
 \be\lb{2dual}
\omega_{ab}=\frac{1}{2}c_{abcd}\omega_{cd},
\ee
or, equivalently,
\be\lb{3dual} c_{abc}\omega_{bc}=0. \ee

In \cite{Bilal} a clear proof that the definition (\ref{2dual}) implies
(\ref{7duality}) was given. It is based in the following identity for the
octonion constants
\be\lb{1id}
c_{abcp}c_{defp}=-3c_{ab[de}\delta_{e]b}-2c_{def[a}\delta_{b]c}+
6\delta^{[d}_{a}\delta^{e}_{b}\delta^{f]}_{c}-2c_{def[a}\delta_{b]c}.
\ee

If (\ref{2dual}) holds it is clear that
$d\omega^{a}_{b}$ will be self-dual. The self-duality of
$\omega^{a}_{c}\wedge\omega^{c}_{b}$ follows using that
$$
\frac{1}{2}c_{abcd}\omega^{c}_{e}\wedge\omega^{e}_{d}=
\frac{1}{4}c_{abcd}c_{edfg}\omega^{c}_{e}\wedge\omega^{f}_{g}
$$
\be\lb{prueba}
=-\frac{1}{2}c_{abfe}\omega^{f}_{c}\wedge\omega^{c}_{e}
+\frac{1}{4}c_{aefg}\omega^{b}_{e}\wedge\omega^{f}_{g}
-\frac{1}{4}c_{befg}\omega^{a}_{e}\wedge\omega^{f}_{g}
+\omega^{a}_{c}\wedge\omega^{c}_{b}
\ee
$$
=-\frac{1}{2}c_{abcd}\omega^{c}_{e}\wedge\omega^{e}_{d}
+2\omega^{a}_{c}\wedge\omega^{c}_{b}\;.
$$
The identity (\ref{1id}) has been used here. From (\ref{prueba}) is
seen that
$$
\frac{1}{2}c_{abcd}\omega^{c}_{e}\wedge\omega^{e}_{d}=
\omega^{a}_{c}\wedge\omega^{c}_{b}
$$
which implies the self-duality of R. The definition (\ref{2dual}) implies (\ref{7duality}),
but the converse is not necessarily true.

\section{The holonomy group of the metric obtained.}

  The purpose of this section is to show that seven dimensional
self-dual manifolds has restricted holonomy $G_2$. Holonomy is the
process of assigning to each closed curve of a manifold the
linear transformation that measures the rotation
resulting when a spinor or vector field is parallel transported
around the given curve. The set of holonomy matrices constitutes a
group, the holonomy group of the manifold. If it is considered only
those curves which are contractible to a point it is
the restricted holonomy. In simply connected manifolds
both groups coincides.

   From the Berger list \cite{Berger} it follows that if a
seven dimensional manifold admits only one covariantly constant spinor
it will have $G_2$ restricted holonomy. It will be shown
now that the seven dimensional self-dual manifolds admits exactly one. The
covariant derivative of a spinor $\eta$ is
\be\lb{covder}
D_i\eta=(\partial_i-\frac{1}{4}\omega_{iab}\gamma^{ab})\eta\;.
\ee

Choosing its components as
\be\lb{espinor}
\eta_{\alpha}=\delta_{\alpha 8}
\ee
it is obtained
$$
D_i\eta=-\frac{1}{4}\omega_{iab}\gamma^{ab}\eta.
$$
  One of the possible representations of the $SO(7)$ gamma
matrices is the antisymmetric and imaginary given by
$$
\gamma^{a}_{\alpha\beta}= i (c'_{a \alpha \beta} +
\delta_{a\alpha}\delta_{8\beta}+\delta_{a\beta}\delta_{8\alpha}),
$$

where the constants $c'_{a \alpha \beta}$ are zero if $\alpha$ or
$\beta$ are equal to $8$, and the octonion constants in other
case. In this representation $(\gamma^{ab}\eta)_{\alpha}=c_{abc}$.
Using (\ref{3dual}) it follows that
$$
D_i\eta=-\frac{1}{4}\omega_{iab}\gamma^{ab}\eta=-\frac{1}{4}\omega_{iab}c_{abc}=0.
$$

With this election for the gamma matrices, this is the only spinor that
satisfies the last equation. But the result must be independent of the
representation, which implies that the seven dimensional self-dual manifolds admits
exactly one covariantly constant spinor and their holonomy is $G_2$.

  The change of an arbitrary field $\Psi$ under infinitesimal parallel transport
is
 \be\lb{paralell}
 \delta\Psi=G_{ab}\delta A^{ab}\Psi\;,
 \ee
where $\delta A^{ab}$ is an infinitesimal area element spanned by
the closed curve taking into consideration.
$G_{ab}=R_{abcd}\Gamma^{cd}$ generate the infinitesimal holonomy
group, being $\Gamma^{cd}$ the generators of $SO(m)$ in the
representation of the field $\Psi$. The restricted holonomy
can be larger than the infinitesimal one.

  It has been mentioned in the introduction that for $G_2$ manifolds it is
possible to construct a $G_2$ invariant closed and co-closed three form.
For this reason it is needed to check that this holds for seven dimensional
manifolds with self-dual spin connection. The most natural candidates
to consider are the $G_2$ equivariant $3$-form $\Phi$ (\ref{3form}) and
its dual $\ast\Phi$ \cite{Bilal}
\be\lb{24form}
 \Phi=\frac{1}{3!}c_{abc}e^a\wedge e^b\wedge
e^c\;,\;\;\; \ast\Phi=\frac{1}{4!}c_{abcd}e^a\wedge e^b\wedge
e^c\wedge e^d. \ee
Taking into account the identity
$$
c_{abp}c_{pcde}=3c_{a[cd}\delta_{e]b}-2c_{b[cd}\delta_{e]a}
$$
it is obtained
$$
d\Phi=-\frac{1}{3!2}c_{abc}e^a\wedge e^b\wedge\omega^{cd}\wedge
e^d= -\frac{1}{3!}c_{abc}c_{cdef}e^a\wedge e^b\wedge
\omega^{ef}\wedge e^d
$$
$$
=\frac{1}{3!}c_{ade}e^a\wedge e^d \wedge \omega_{eb} \wedge
e^b=-2d\Phi\;.
$$
From here follows \be\lb{clauphi} d\Phi=0\;. \ee So, $\Phi$ is a
closed form. Similarly, using (\ref{1id}) follows that
$$
d\ast\Phi=-\frac{1}{4!6}c_{abcd}\omega_{ae}\wedge e^e \wedge e^b
\wedge e^c\wedge e^d
=-\frac{1}{4!12}c_{aefg}c_{abcd}\omega^{fg}\wedge e^e \wedge e^b
\wedge e^c \wedge e^d
$$
$$
=\frac{1}{4!3}c_{fbcd}\omega^{fe}\wedge e^e \wedge e^b\wedge
e^c\wedge e^d= -2d\ast\Phi\;,
$$
which implies the closure of $\ast\Phi$.

\section{The Bryant-Salamon construction}

  A construction of $G_2$ manifolds starting with four dimensional quaternionic Kahler
manifold as a base space has been given by Bryant and Salamon in \cite{Salamon} and reconsidered recently
in \cite{Mahapatrah}. Those spaces are constructe as $R^3$ bundles over a quaternionic base.
In the hyperkahler limit the bundle will be trivial, i.e, it will be the
global product of the hyperkahler one with $R^3$.

    Before describe this construction it is convenient to review certain propierties of
quaternionic spaces. By definition \cite{Fre} a quaternionic Kahler space is a Riemannian space
of real dimension $4N$ endowed with a metric
$$
ds^2=g_{\mu\nu}(x)dx^{\mu}dx^{\nu}
$$
and a set of three complex structures $J^{i
\beta}_{\alpha}$ satisfying the quaternionic algebra
\be\lb{almcomp} J^{i \beta}_{\alpha}J^{j
\gamma}_{\beta}=-\delta_{ij}\delta_{\alpha}^{\gamma}+\epsilon_{ijk}J^{k
\gamma}_{\alpha}.
\ee
 The metric is quaternionic
hermitian:
$$
g(J^iX,J^iY)=g(X,Y) ,
$$
from where follows that $J^{i}_{\beta\alpha}=-J^{i}_{\alpha\beta}$.
The holonomy group $H \subset Sp(n)\times Sp(1)$, and if the manifold has
scalar curvature equal to zero it will be called hyperkahler.
From the complex structures $J^{i}$ is possible to construct
the hyperkahler triplet of $2$-forms given by
$$
\Omega^i=\Omega^{i}_{\mu\nu}dx^{\mu}dx^{\nu};\;\;\;
\Omega^{i}_{\mu\nu}=g_{\mu\omega}(J^i)^{\omega}_{\nu}.
$$
and the three local $1$-forms
\be\lb{onefor}
A^{i}=\omega^{mn}J^{i}_{mn}
\ee
where $\omega^{mn}$ represents the antiself-dual part of the spin connection.
The hyperkahler form is covarianly closed with respect to the connection $A^i$
; namely
$$
\nabla_{\alpha}\Omega^{i}=d\Omega^{i}+\epsilon_{ijk}A^j \wedge \Omega^{k}=0.
$$
Defining the $SU(2)$ curvature
$$
F^i=dA^i+\epsilon_{ijk}A^j \wedge A^k,
$$
it follows that
\be\lb{rela}
F^i=\kappa \Omega^i.
\ee
Here $\kappa$ denotes the scalar curvature. Any quaternionic metric is an
Einstein space with curvature $\kappa$ and
$$
R_{mn}=3\kappa g_{mn}.
$$
It is important for the purposes of the present work to remark that in four
dimensions a quaternionic-Kahler metric is an Einstein metric with self-dual Weyl
tensor
$$
W_{ab}=\frac{c_{abcd}}{2}W_{cd}.
$$
In the hyperkahler case $\kappa=0$, so $F^i$ will be flat and the manifold
will be Ricci-flat.

   The Bryant Salamon result is that the following
\be\lb{anzatz}
ds^2=\frac{1}{\sqrt{2\kappa|u|^2+c}}(du^i+\epsilon^{ijk}A^ju^k)^2+\sqrt{2\kappa|u|^2+c}ds^2_4.
\ee
is a $G_2$ holonomy metric. An straightforward proof can be found in the
reference \cite{Mahapatrah}, there is shown that (\ref{anzatz}) has a self dual spin
connection and in consecuence the restricted holonomy will be $G_2$.
The metric $ds^2_4$ corresponds to the quaternionic base space; the total one
is topologically an $R^3$ bundle with coordinates $u_i$, and $c$ is an integration constant.

   In the hyperkahler limit $\kappa=0$ it is possible to gauge away the connection $A^i$;
the resulting manifold is the trivial product of
$R^3$ by the hyperkahler manifold, which is non-compact and with holonomy $G_2$.
Non-compact backgrounds are of interest in M-theory as well \cite{Mahapatrah}.
But it should be noted that a manifold that is globally the cartesian product
of $R^3$ with any four dimensional manifold with self-dual spin connection
will be a $G_2$ holonomy space. The reason is the following: the spin connection
obviously will be independent on the coordinates of $R^3$ and the condition
(\ref{2dual}) will reduce to the ordinary self-duality condition in four dimensions
; it follows that if the four dimensional base space is self-dual the restricted
holonomy will be $G_2$.

  Expressing the $R^3$ part of (\ref{anzatz}) in polar coordinates it is obtained the following expression
$$
ds^2=\frac{dr^2}{(1-\frac{4c}{r^4})}+\frac{r^2}{4\kappa}(1-\frac{4c}{r^4})g_{ab}
(dx^a+\xi^a_i A^i)(dx^b+\xi^b_j A^j)+\frac{r^2}{2}ds_4^2,
$$
where $g_{ab}$ and $\xi$ are the metric and the killing vectors of $S^2$ respectively. In this
coordinate system is more clear that the metric is asymptotically a cone; in the limit $r\rightarrow\infty$
it is seen that
$$
ds^2\sim dr^2+r^2d\Omega
$$
where the part related with $\Omega$ is independent of $r$. In other words, for large $r$ (\ref{anzatz}) is a
cone over a six dimensional space constructed as an $S^2$ bundle over a quaternionic kahler space with the metric
$ds^2_4$. This six dimensional manifold has "weak" holonomy $SU(3)$ \cite{Gibb2};
such manifolds are of great interest in compactification of the type IIA supergravity.

\section{Toric quaternionic geometry in four dimensions.}
\subsection{The Ward and the Calderbank-Pedersen descriptions }
   In the last subsection it has been described how to construct $G_2$ holonomy spaces
starting with hyperkahler and quaternionic kahler ones.
The aim of this work is to investigate the case in which there are
two commuting isometries. The hyperkahler case was discussed in \cite{Calderbank2}
and is related with the monopole solution appearing in the context of the $(2+1)$
Einstein gravity \cite{Ward}.

    The main result needed here is that the four dimensional euclidean metric
\be\lb{sdvacum}
ds^2=V_{\eta}(d\rho^2+d\eta^2+\rho^2d\phi^2)+V^{-1}_{\eta}(d\psi+\rho V_{\rho}d\phi)^2
\ee
will be hyperkahler if and only if the function $V$ is an Axisymmetric
Harmonic Function (AHF), i.e, if it satisfies the monopole equation
\be\lb{monopole}
V_{\eta\eta}+\rho^{-1}(\rho V_{\rho})_{\rho}=0.
\ee

The metric (\ref{sdvacum}) has an $U(1)\times U(1)$ isometry because
the two killing spinors $\frac{\partial}{\partial \phi}$ and
$\frac{\partial}{\partial \psi}$ commutes.

   Many properties of the AHF are well know.
It has been shown \cite{Ward} that any solution V of the equation
$$
V_{\eta\eta}-\rho^{-1}(\rho V_{\rho})_{\rho}=0
$$
can be expressed in integral form as
\be\lb{intexp}
V(\eta, \rho)=\frac{1}{2\pi}\int^{2\pi}_0 G(\rho sen(\theta)+\eta)d\theta.
\ee

Here $G(x)$ denotes an arbitrary function of one variable. For a given solution
$V(\eta, \rho)$ the function $V(i\eta, \rho)$ will be
a solution of (\ref{monopole}), and $W(\eta, \rho)=V(i\eta, \rho)+ V(-i\eta, \rho)$
will be a real AHF. This integral representation will be used in the following sections.

   In a recent work Calderbank and Pedersen have given a complete description
of the four dimensional non-Ricci flat Einstein metrics with self-dual Weyl
tensor and two commuting isometries, in terms of certain eigenfunctions of the hyperbolic laplacian in two dimensions
\cite{Pedersen}. This case is of interest as well, because such metrics will be
quaternionic-kahler.

     Their statement is that for any Einstein-metric with
self-dual Weyl tensor and nonzero scalar curvature possessing two linearly
independent commuting Killing fields there exists a coordinate system in which
the metric g has locally the form
$$
g=\frac{F^2-4\rho^2(F^2_{\rho}+F^2_{\eta})}{4F^2}\frac{d\rho^2+d\eta^2}{\rho^2}
$$
\be\lb{metric}
+\frac{[(F-2\alpha F_{\rho})\alpha-2\rho F_{\eta}\beta]^2+[-2\rho F_{\eta}\alpha
+(F+2\rho F_{\rho})\beta]^2}{F^2[F^2-4\rho^2(F^2_{\rho}+F^2_{\eta})]},
\ee
where $\alpha=\sqrt{\rho}d\phi$ and $\beta=(d\psi+\eta d\phi)/\sqrt{\rho}$
and $F(\rho, \eta)$ is a solution of the equation
\be\lb{backly}
 F_{\rho\rho}+F_{\eta\eta}=\frac{3F}{4\rho^2}.
\ee

on some open subset of the half-space $\rho>0$. On the open set defined by $F^2 > 4\rho^2(F^2_{\rho}+F^2_{\eta})$ g has positive scalar curvature,
whereas $F^2 < 4\rho^2(F^2_{\rho}+F^2_{\eta})$ -g is self-dual with negative
scalar curvature.

  It follows that quaternionic metrics with torus symmetry has positive signature.
The three $1$-forms (\ref{onefor}) have a remarkable simple expression in terms of F,
\be\lb{threefor}
A^1= \frac{1}{F}[-\rho F_{\eta}\frac{d\rho}{\rho}+(\frac{1}{2}F+\rho F_{\rho})\frac{d\eta}{\rho}],\;\;
A^2=\frac{\alpha}{F},\;\; A^3=\frac{\beta}{F}.
\ee
and the relation (\ref{rela}) holds with $\kappa=1$.

   Both statements presented here can be interpreted as methods to construct
quaternionic kahler and hyperkahler geometries starting with solutions of
two linear equations of second order, namely (\ref{monopole}) and (\ref{backly}).
As we will see, this equations are related by a Backglund transformation,
which implies any solution of one of them allow us to construct a solution
of the other one. This fact can be exploited to find examples of quaternionic
metrics.

\subsection{The Backglund transformation.}

  To prove that (\ref{monopole}) and (\ref{backly}) are Backglund transformed it is needed to introduce
the Joyce system of equations \cite{Pedersen}
\be\lb{system}
(S_0)_{\rho}+(S_1)_{\eta}=S_0/\rho,\;\;\;\; (S_0)_{\eta}-(S_1)_{\rho}=0
\ee
where $S_0$ and $S_1$ are unknown functions. Selecting $S_0=H_{\rho}$ and $S_1=H_{\eta}$ the second equation
will be trivial and the first became $H_{\rho\rho}+H_{\eta\eta}=H_{\rho}/\rho$ ($H$ is usually
called a Tod coordinate). Now, taking $H=\rho^{1/2}F$ it follows that F satisfies (\ref{backly}). Conversely,
selecting $S_0=-\rho V_{\eta}$ and $S_1=\rho V_{\rho}$ the first equation is trivial and the
second one is $V_{\eta\eta}+\rho^{-1}(\rho V_{\rho})_{\rho}=0$, in other words V is an AHF.

    It is seen that a solution $V$ of (\ref{monopole}) allows construct a solution F of
(\ref{backly}) integrating the system
\be\lb{shu}
H_{\rho}=-\rho V_{\eta};\;\;\;\; H_{\eta}=\rho V_{\rho}
\ee
and defining $F=H/\rho^{1/2}$. Conversely, a solution F of (\ref{backly}) allows to construct
a solution V of (\ref{monopole}) defining $H=\rho^{1/2}F$ and integrating (\ref{shu}). By construction
the equation for F is the integrability condition for V and viceversa; the relation
between them is an example of a Backglund mapping. This
gives a method to construct quaternionic-kahler metrics starting with an hyperkahler example and
viceversa, in both cases there is a torus symmetry. The representation
(\ref{intexp}) together with (\ref{shu}) can be exploited to find non trivial examples
of quaternionic kahler spaces, with only selecting an arbitrary function of one
variable. This is the purpose of the next subsection.

\subsection{Examples of quaternionic spaces with torus symmetry.}

The following are quaternionic metrics constructed starting with an
arbitrary function G(x).

         A. The trivial four dimensional toric metric
$$
ds^2=d\rho^2+d\eta^2+\rho^2d\phi^2+d\psi^2
$$
corresponds to the monopole
$$
V=\eta
$$
(This AHF holds using $G(x)=x Log(x)$ in the integral expression (\ref{intexp})).
The equations (\ref{shu}) are in this case
$$
H_{\rho}=-\rho;\;\;\;\;  H_{\eta}=0,
$$
and the eigenfunction F is given by
$$
F=\frac{\rho^{3/2}}{2}.
$$
The insertion of the last eigenfunction in (\ref{metric}) gives the following metric:
\be\lb{quaterna}
g=-\frac{2d\rho^2+2d\eta^2}{\rho^2}-\frac{\rho^2(1+3\rho)+\eta^2(9+7\rho)}{2\rho^5}d\phi^2
-\frac{8}{\rho^4}d\psi^2-\frac{16\eta}{\rho^4}d\phi d\psi.
\ee
The inequality $F^2<4\rho^2(F^2_{\rho}+F^2_{\eta})$ holds for $\rho>0$.
Invoking the 'alderbank-Pedersen theorem, we see that for $\rho>0$ the quaternionic kahler metric
is -g. In this case $\kappa=-1$ and the three
1-forms $A^i$ are
$$
A^1=\frac{2 d\eta}{\rho};\;\;\; A^2=\frac{2d\phi}{\rho};\;\;\; A^3=\frac{2\eta}{\rho^2}d\phi+ \frac{2}{\rho^2}d\psi.
$$
The metric constructed in this example is defined for all the positive values of $\rho$.

        B. With the function $G(x)=Log(x)x^3$ it is found the monopole
$$
V=3\eta\rho^2-2\eta^3
$$
and the hyperkahler metric
$$
ds^2=(3\rho^2-6\eta^2)(d\rho^2+d\eta^2+\rho^2d\phi^2)
+\frac{1}{3\rho^2-6\eta^2}(d\psi+6\eta\rho^2 d\phi)^2.
$$
The Backglund transformed F results
$$
F=\frac{3}{4}\rho^{3/2}(4\eta^2-\rho^2)
$$
and the corresponding metric is
\be\lb{quaternb}
g=g_{\rho\rho}(d\rho^2+d\eta^2)+g_{\phi\phi}d\phi^2+g_{\psi\psi}d\psi^2+2 g_{\phi\psi}d\phi d\psi,
\ee
where the components of the metric tensor are
$$
g_{\rho\rho}=-\frac{4(8\eta^4+6\eta^2\rho^2+3\rho^4)}{(\rho^3-4\eta^2\rho)^2}
$$
$$
g_{\phi\phi}=-\frac{8\eta^4\rho^2(19+5\rho)+16\eta^6(9+7\rho)+\rho^6(1+35\rho)
+3\eta^2\rho^4(35+61\rho)}{9\rho^5(\rho^2-4\eta^2)^2(8\eta^4+6\eta^2\rho^2+3\rho^4)}
$$
$$
g_{\psi\psi}=-\frac{64(4\eta^4+\rho^4)}{9\rho^4(\rho^2-4\eta^2)^2(8\eta^4+6\eta^2\rho^2+3\rho^4)}
$$
$$
g_{\psi\phi}=-\frac{32(8\eta^5-\rho^2\eta^3+3\eta\rho^4)}{9\rho^4(\rho^2-4\eta^2)^2(8\eta^4+6\eta^2\rho^2+3\rho^4)}.
$$

As in the example A, $F^2<4\rho^2(F^2_{\rho}+F^2_{\eta})$ for $\rho>0$, -g is a quaternionic kahler
metric and $\kappa=-1$. The three forms $A^i$ are given by
$$
A^1=\frac{8\eta}{\rho^2-4\eta^2}d\rho+\frac{4(\rho^2-2\eta^2)}{\rho(\rho^2-4\eta^2)}d\eta
$$
$$
A^2=\frac{4}{3\rho(\rho^2-4\eta^2)}d\phi; \;\;\;\
A^3=\frac{4\eta}{3\rho(\rho^2-4\eta^2)}d\phi+\frac{4}{3\rho(\rho^2-4\eta^2)}d\psi.
$$

The metric (\ref{quaternb}) is singular at $\rho\rightarrow 0$ and at the
lines $2|\rho|=|\eta|$.

       C. The powers $G(x)=x^n$ and $G(x)=Log(x)x^{2n+1}$ can be integrated out giving polynomial solutions
of higher degree. For instance $G(x)=Log(x)x^5$ gives
$$
V=3\eta \rho^2-2\eta^3;\;\;\;\;F=\frac{1}{\eta}(8\eta^4+ 40 \eta^2 \rho^2+15 \rho^4).
$$
The even powers $G(x)=x^{2n}Log(x)$ can be integrated too, but the expressions of the
metrics are more complicated by the appearance of logarithm terms. For example, with
$G(x)=Log(x)x^{2}$ it is obtained
$$
V=6\eta^2-\rho^2-6\eta^2\sqrt{1+\frac{\rho^2}{\eta^2}}+2(\rho^2-2\eta^2)Log(\frac{2}{\eta+\sqrt{\eta^2+\rho^2}})
$$
$$
F= \frac{4}{3}\eta^3 - 4\eta \rho^2 - \frac{4}{3}\eta^2(\eta^2 + \rho^2)+
\frac{8}{3} \rho^2(\eta^2 + \rho^2) + 4\eta \rho^2 Log(\frac{2}{\eta + \sqrt{\eta^2 + \rho^2)}}).
$$

The expression for the quaternionic metric and the hyperkahler
one corresponding to such solutions is very large and difficult to simplify.

           D. The function $G(x)=e^{x}$ gives
$$
V=e^{\pm i\eta} I_0(\rho)+ c.c
$$
where $I_n(\rho)$ denotes the modified Bessell function of the first kind, which are solutions
of the equation
$$
\rho^2 H''(\rho)+\rho H'(\rho)-(\rho^2+ n^2)H(\rho)=0.
$$
 The hyperkahler space that corresponds to this monopole is:
$$
ds^2=\rho I_1(\rho)(d\rho^2+d\eta^2+\rho^2d\phi^2)cos(\eta)
$$
\be\lb{kahlermia}
+\frac{1}{\rho I_1(\rho)cos(\eta)}
\{d\psi+\rho[I_1(\rho)+\frac{\rho}{2}(I_0(\rho)+I_2(\rho))]sin(\eta)d\phi\}^2.
\ee
The Backglund transformed eigenfunction F is given by
$$
F= \sqrt{\rho}I_1(\rho)e^{\pm i\eta}+ c.c.
$$
and from this solutions it follows the quaternionic metric
$$
ds_{qk}=\frac{\Theta(\rho, \eta)}{4\rho I_1(\rho)^2}(d\rho^2+d\eta^2)
+\frac{[2\rho^{3/2}I_1(\rho)\beta cos(\eta)-\rho^{3/2} (I_0(\rho)+ I_2(\rho))\alpha sin(\eta)]^2}{\Phi(\rho,\eta)}
$$
\be\lb{metricmia}
+\frac{[2\rho^{3/2}I_1(\rho)\alpha cos(\eta) + \sqrt{\rho}(\rho I_0(\rho)+ 2 I_1(\rho)+ \rho I_2(\rho))
\beta sin(\eta)]^2}{\Phi(\rho,\eta)},
\ee
where it has been defined
$$
\Theta(\rho, \eta)=\rho I_0(\rho)^2 + 2 I_1(\rho)I_2(\rho)+ \rho I_2(\rho)^2+ 4\rho I_1(\rho)^2 ctg(\eta)^2+
2 I_0(\rho)(I_1(\rho)+\rho I_2(\rho))
$$
and
$$
\Phi(\rho,\eta)=-\rho^2 I_1(\rho)^2 sin(\eta)^2[4\rho^2 I_1(\rho)^2 cos(\eta)^2- I_1(\rho)^2 sin(\eta)^2+
(\rho I_0(\rho)+ I_1(\rho)+ \rho I_2(\rho))^2 sin(\eta)^2].
$$
It will be quaternionic for the regions of the plane $(\rho, \eta)$ in which $F^2< 4\rho^2(F^2_{\rho}+F^2_{\eta})$.
In those regions $k=-1$. The three one forms $A^i$ are
$$
A^1=\frac{1}{2}[1 + tg(\eta)+ \frac{\rho}{I_1(\rho)}tg(\eta)(I_0(\rho)+I_2(\rho))]
\frac{d\eta}{\rho}-tg(\eta)d\rho
$$
$$
A^2=\frac{\alpha}{\sqrt{\rho}I_1(\rho)cos(\eta)};\;\;\;\; A^3=\frac{\beta}{\sqrt{\rho}I_1(\rho)cos(\eta)}.
$$
The radial component of the metric (\ref{metricmia}) shows that some of the singularities are
the zeros of $I_1(\rho)$.

\subsection{The m-pole solutions}

 In the previous subsection it have been constructed quaternionic-kahler
metrics starting with an arbitrary AHF and solving the Backglund equations.
By completeness it will be discussed here the spaces corresponding to the
$m$-pole solutions investigated in \cite{Dancer} and \cite{Pedersen}.
It has been analyzed by Anguelova and Lazaroiu the dynamics of the M-theory
on toric $G_2$ cones constructed with $m$-pole spaces as base manifolds
in \cite{Angelova} and \cite{Angelito}. Such examples leads to toric Einstein
self-dual spaces of positive scalar curvature which are complete (thus compact)
and admiting only orbifold singularities. As such,
they seem to be the best candidates with $U(1)\times U(1)$ isometry
for which the physical analysis given in
\cite{Witten2} can be applied.
In the first reference of \cite{Pedersen} it has been described the moduli
space corresponding to the $3$-pole solutions and has been shown that they encode
some well known examples appearing in the physics, like the Bianchi type spaces.
It will be shown that the hyperkahler metrics corresponding to the $3$-pole
solution are those discussed in \cite{Calderbank2} which gives rise to the
$3$-dimensional Eguchi-Hanson like Einstein-Weyl metrics after the quotient by
one of the isometries. The following exposition follows closely those given in the
references \cite{Pedersen}.

   The basic eigenfunctions F of (\ref{backly}) which we need to consider are
\be\lb{solu}
F(\rho, \eta, y)=\frac{\sqrt{(\rho)^2+(\eta-y)^2}}{\sqrt{\rho}}
\ee

where the parameter $y$ takes arbitrary real values. Using the Backglund transformation it is found
the basic monopole
\be\lb{soludos}
V(\eta, \rho, y)=-Log[\eta-y + \sqrt{\rho^2 + (\eta-y)^2}].
\ee

   Being the equations for F and V linear, for any set of
real numbers $w_i$ the functions
\be\lb{superpos}
F=\sum^{k+1}_{j=0} w_i F(\rho, \eta, y_j).
\ee
\be\lb{super2}
V=\sum^{k+1}_{j=0} w_i V(\rho, \eta, y_j)
\ee
will be solutions too. For this reason the $2$-pole functions
given by
$$
F_1=\frac{1+\sqrt{\rho^2+\eta^2}}{\sqrt{\rho}};\;\;\; F_2=\frac{\sqrt{(\rho)^2+(\eta+1)^2}}{\sqrt{\rho}}-
\frac{\sqrt{(\rho)^2+(\eta-1)^2}}{\sqrt{\rho}},
$$
are eigenfunctions of the hyperbolic laplacian. The first one gives rise to the spherical metric,
while the second one gives rise to the hyperbolic metric
$$
ds^2=(1-r_1^2-r_2^2)^{-2}(dr_1^2+dr^2_2 + r_1^2d\theta_1^2 + r_2^2d\theta_2^2).
$$
The relation between the coordinates $(r_1, r_2)$ and $(\rho, \eta)$ can be extracted from
the relation
$$
(r_1 + i r_2)^2=\frac{\eta-1+i\rho}{\eta+1+i\rho}.
$$
The hyperkahler metrics corresponding to both cases are
\be\lb{hyper1}
ds^2=-\frac{1}{\sqrt{\rho^2 + \eta^2}}(d\rho^2+d\eta^2+\rho^2d\phi^2)
-\sqrt{\rho^2 + \eta^2}(d\psi+ \frac{\eta}{\sqrt{\rho^2 + \eta^2}}d\phi)^2,
\ee
and
$$
ds^2= \frac{\sqrt{\rho^2+(\eta-1)^2}-\sqrt{\rho^2+(\eta+1)^2}}{\sqrt{\rho^2+(\eta+1)^2}\sqrt{\rho^2+(\eta-1)^2}}
(d\rho^2+d\eta^2+\rho^2d\phi^2)
$$
\be\lb{hyper2}
+\frac{\sqrt{\rho^2+(\eta+1)^2}\sqrt{\rho^2+(\eta-1)^2}}{\sqrt{\rho^2+(\eta-1)^2}-\sqrt{\rho^2+(\eta+1)^2}}
[d\psi+(\frac{\eta+1}{\sqrt{\rho^2+(\eta+1)^2}}-\frac{\eta-1}{\sqrt{\rho^2+(\eta-1)^2}})d\phi]^2.
\ee

    The general "$3$-pole" solutions are
$$
F=\frac{1}{\sqrt{\rho}}+\frac{b+c/m}{2}\frac{\sqrt{\rho^2+(\eta+m)^2}}{\sqrt{\rho}}
+\frac{b-c/m}{2}\frac{\sqrt{\rho^2+(\eta-m)^2}}{\sqrt{\rho}}.
$$
By definition $-m^2=\pm 1$, which means that $m$ can be imaginary or real.
The corresponding solutions are denominated type I and type II respectively.
It is convenient to introduce the Eguchi-Hanson like coordinate system defined by
$$
\rho= \sqrt{R^2 \pm 1} cos(\theta), \;\;\;\; \eta=R sin(\theta),
$$
where $\theta$ takes values in the interval $(-\pi/2, \pi/2)$. In this coordinates
\be\lb{unade}
\sqrt{\rho}F=1+b R+c sin(\theta),
\ee
\be\lb{dosde}
\rho^{-1}[\frac{1}{4}F^2-\rho^2(F_{\rho}^2+F_{\eta}^2)]=
\frac{b(R \mp b) + c(sin(\theta) + c)}{R^2 \pm sin^2(\theta)}.
\ee
and the family of self-dual metrics corresponding to the $3$-pole are expressed as
$$
ds^2=\frac{b^2-c^2+(bR-cS)}{(1+b R+c S)^2}(\frac{dR^2}{R^2-1}+\frac{dS^2}{1-S^2})
$$
$$
\frac{1}{(1+b R+ cS)^2(b^2-c^2+(bR-cS))(R^2-S^2)}
$$
$$
*((R^2-1)(1-S^2)((bR-cS)d\varphi + (cR-bS) d\psi)^2
$$
\be\lb{enclo}
+ ((b(R^2-1)S+c(1-S^2)R)d\varphi + (c(R^2-1)S + b(1-S^2)R + (R^2-S^2)d\psi)^2)
\ee
It has been denoted $S= sin(\theta)$ here. The expression (\ref{enclo}) includes some
well known metrics. Let us focus in the type I case.
The formulas (\ref{unade}) and (\ref{dosde}) allows to determine the domain of
definition of the metric (\ref{metric}). When b is nonzero for a given value of $\theta$, $F=0$ if
$R=-(1+c sin(\theta))/b$  and  $(\frac{1}{4}F^2-\rho^2(F_{\rho}^2+F_{\eta}^2))=0$
if $R=(b^2+c^2+c sin(\theta))/b$. The case $c=0$ correspond to a bi-axial
Bianchi IX metric \cite{Pedor}. The domains of definition are $(-\infty, R_{\infty}),
(R_{\infty}, R_{\pm})$, and $(R_{\pm}, \infty)$. In the first two cases the curvature is negative,
and in the last one positive, and in the two last cases there is an unremovable singularity at
$R=R_{\pm}$. In the case $b=0$ for $c>1$ and $c<1$ the metric will be of
Bianchi VIII type \cite{Caldo}. The case $c=1$ corresponds to the Bergmann metric on $CH^2$.

  For the type II case, the range of R is $(1,\infty)$ but the moduli space is more complex
that in the type I case. For the lines $b=\pm c$ it is obtained the hyperbolic metric if $b<0$
and the spherical metric if $b>0$. If $(b,c)=(1,0)$ it is obtained the Fubbini-Study metric on
$CP^2$ whereas the points $(0,1), (-1,0)$ and $(0,-1)$ yield again the Bergmann metric on $CH^2$.
Along the lines joining $(1,0)$ with others we have bi-axial Bianchi metric IX, while along the lines
between $(0,1), (-1,0)$ and $(0,-1)$ the metric is Bianchi VIII. A more complete description
is given in \cite{Pedersen}.

   The triplet of one forms corresponding to this family of metrics is
$$
A^1=A^1_{+} +A^1_{-},
$$
$$
A^2=\frac{\sqrt{(R^2 \pm 1)(1-S^2)}}{(1+bR+cS)}d\phi\;\;\;
A^3=\frac{d\psi+\eta d\phi}{(1+bR+cS)},
$$
where it has been defined
$$
A^1_{\pm}=A^1_{1\pm}+A^1_{2\pm}+A^1_{3\pm},
$$
with
$$
A^1_{1\pm}=\frac{(b \pm c/m)(S R \pm m)}{(1+bR+cS)\sqrt{(1 - S^2)(R^2\pm 1)+ ( S R\pm m)^2}}*
$$
$$
(\frac{S\sqrt{R^2\pm1}}{2\sqrt{1-S^2}}dS
+ \frac{R \sqrt{1-S^2}}{\sqrt{R^2\pm 1}} dR )
$$
and
$$
A^1_{2\pm}=\frac{(b \pm c/m)\sqrt{(1 - S^2)(R^2\pm 1)
+ ( S R\pm m)^2}}{2(1+bR+cS)\sqrt{R^2 \pm 1}}
(R\frac{dS}{\sqrt{1-S^2}}+ S dR)
$$
and
$$
A^1_{3\pm}=[\frac{(b \pm c/m)(R^2\pm 1)^{1/2}(1 - S^2)^{1/2}}{2\sqrt{(1 - S^2)(R^2\pm 1)+ ( S R \pm m)^2}}-
\frac{(b \pm c/m)\sqrt{(1 - S^2)(R^2\pm 1)+ (S R \pm m)^2}}{4(1 - S^2)^{1/2}(R^2\pm 1)^{1/2}}]*
$$
$$
\frac{\sqrt{(1 - S^2)(R^2\pm 1)}}{(1+bR+cS)\sqrt{R^2 \pm 1}}
(R\frac{dS}{\sqrt{1-S^2}}+ S dR)
$$

(The sign $\pm$ in $(R^2\pm 1)$ depends only on the metric in consideration,
it is $+$ for type I and $-$ for type II.)

  The Backglund transformed function V reads
$$
V=Log(\rho)+\frac{(b+c/m)}{2}Log[\frac{\eta-m+\sqrt{(\eta-m)^2+\rho^2}}{\rho}]
$$
$$
+\frac{(b-c/m)}{2}Log[\frac{\eta+m+\sqrt{(\eta+m)^2+\rho^2}}{\rho}].
$$
For the type I case this is the potential for an axially symmetric circle of charge, while the
type II case corresponds to two point sources on the axis of symmetry. The hyperkahler metrics
obtained are encoded in the following expression
$$
ds^2=\frac{bR+c\sqrt{1-S^2}}{R^2\pm (1-S^2)}
(d\rho^2+d\eta^2+\rho^2d\phi^2)
$$
\be\lb{hyper2}
+\frac{R^2 \pm (1-S^2)}{bR+c\sqrt{1-S^2}}
[ d\psi +\frac{R^2\pm(1-S^2)-b(R^2\pm 1)\sqrt{1-S^2}
+ c R S^2}{R^2
\pm (1-S^2)}d\phi ]^2.
\ee
This manifolds have been investigated recently in \cite{Calderbank2} and it has been
shown that the quotient of (\ref{hyper2}) with $\frac{\partial}{\partial \phi}$ gives the
Eguchi-Hanson type Einstein-Weyl metrics in D=3.

    The continuum limit of the expressions (\ref{superpos}) and (\ref{super2}) are
\be\lb{continium}
F(\rho, \eta)=\int w(y) F(\rho, \eta, y)dy.
\ee
\be\lb{continium2}
V(\rho, \eta)=\int w(y) V(\rho, \eta, y)dy.
\ee

where $w(y)$ is a distribution with compact support in $R$.
A choice of $w(y)$ for which at least one of the integrals (\ref{continium}) and
(\ref{continium2}) converges gives rise to an smooth solution.
For instance for $w(y)=y/(y^2+1)^2$ it is obtained
the following non-trivial monopole
$$
V(\rho, \eta)=\frac{cos(\frac{1}{2}Arg(1-2i\eta-\eta^2-\rho^2))
Log(\frac{|1-i\eta-\sqrt{(1-i\eta)^2+\rho^2}|}{|1+i\eta-\sqrt{(1-i\eta)^2+\rho^2}|})}{\sqrt{|(1-i\eta)^2+\rho^2}|}
$$
$$
+\frac{sin(\frac{1}{2}Arg(1-2i\eta-\eta^2-\rho^2))
Arg(\frac{1-i\eta-\sqrt{(1-i\eta)^2+\rho^2}}{1+i\eta-\sqrt{(1-i\eta)^2+\rho^2}})}{\sqrt{|(1-i\eta)^2+\rho^2}|},
$$
and from (\ref{sdvacum}) follows an hyperkahler metric. But (\ref{continium}) is divergent for
this distribution.

    To conclude this subsection it should be mentioned that higher $m$-pole solutions
have been considered in \cite{Angelova} and \cite{Angelito}, and that quaternionic spaces with torus symmetry
have been investigated recently in \cite{Valent} using the harmonic
space formalism.

\section{$G_2$ holonomy metrics with torus symmetry and supergravity backgrounds.}

   In this subsection will be constructed the $G_2$ holonomy metrics corresponding to the examples A and B.
After extend them to a vacuum configuration of the eleven dimensional supergravity it will be obtained
type IIA backgrounds by reduction along one of the isometries.

    In \cite{Angelova} it has been found the explicit form of (\ref{anzatz}) when the base space
(and, in consequence, the total one) has torus symmetry. The expression is
\be\lb{quatera}
ds^2=\frac{dr^2}{h(r)}+\frac{r^2}{2}[ U_{\phi\phi}d\phi^2+ U_{\phi\psi}d\phi d\psi+ U_{\psi\psi}d\psi^2+
Q_{\phi}d\phi+Q_{\psi}d\psi+ g_{\rho\rho}(d\eta^2+d\rho^2)+ H ].
\ee
where it has been defined
$$
h(r)=1-4c/r^4
$$
$$
U_{11}=g_{\phi\phi} +h(r)\frac{u_1^2(\rho^2+\eta^2)+(u_2 \eta+u_3 \rho)^2}{2 \rho F^2},
$$
$$
U_{22}=g_{\psi\psi}+h(r)\frac{u_1^2+u_2^2}{2 \rho F^2},
$$
$$
U_{12}=U_{21}=g_{\phi\psi}+h(r)\frac{(u_1^2+u_2^2)\eta+u_2 u_3 \rho}{2 \rho F^2},
$$
$$
Q_{\phi}=h(r)\frac{1}{\sqrt{\rho}F}[u_1(\eta du_2+ \rho du_3)-(u_2 \eta+ u_3 \rho)du_1
-u_1(u_3 \eta-u_2 \rho)A^1],
$$
$$
Q_{\psi}=h(r)[\frac{u_1 du_2-u_2 du_1-u_1 u_3 A^1}{\sqrt{\rho}F}],
$$
$$
H=h(r)[|d\overrightarrow{u}|^2+(u_1^2+u_2^2)(A^1)^2-2 A^1 (u_3 du_2-u_2 du_3)].
$$

The second rank tensor $g_{ab}$ is the metric of the base manifold. The product metric of (\ref{quatera}) with $M^4$
\be\lb{mamas}
ds^2_{11}=ds_{M}^2+\frac{dr^2}{h(r)}+\frac{r^2}{2}[ U_{\phi\phi}d\phi^2+ U_{\phi\psi}d\phi d\psi+ U_{\psi\psi}d\psi^2+
Q_{\phi}d\phi+Q_{\psi}d\psi+ g_{\rho\rho}(d\eta^2+d\rho^2)+ H ].
\ee
is a vacuum configuration of the eleven dimensional supergravity \cite{Duff}. With the help of the quantities
$$
\alpha_1=\frac{U_{22}Q_{\phi}-U_{12}Q_{\psi}}{detU},\;\;\;\alpha_2=\frac{U_{11}Q_{\psi}-U_{12}Q_{\phi}}{detU},
$$
$$
h=H+U_{11}\alpha_{1}^2+2U_{12}\alpha_1\alpha_2+U_{22}\alpha_{2}^2,
$$
$$
\phi_1=\phi,\;\;\;\phi_2=\psi,
$$
the metric (\ref{mamas}) is expressed in more simple manner as
$$
ds^2_{11}=ds_{M^4}^2+\frac{dr^2}{h(r)}+\frac{r^2}{2}[ U_{ij}(d\phi_i+\alpha_i)(d\phi_j+\alpha_j)+ h].
$$
The last expression takes the usual form of the Kaluza-Klein anzatz
\be\lb{colo}
ds^2_{11}=e^{-\frac{2}{3}\varphi_{D}}G_{\mu\nu}dx^{\nu}dx^{\mu}+e^{\frac{4}{3}\varphi_{D}}(d\phi+dx^{\mu}C_{\mu}(x))^2.
\ee
with the dilaton field and the RR 1-form defined by
$$
\varphi_D=\frac{3}{4}Log(\frac{r^2U_{11}}{2}),
$$
$$
C=\frac{U_{12}d\psi+Q_{\phi}}{U_{11}}.
$$
The reduction of (\ref{colo}) along $\phi_1$ gives the following IIA metric:
\be\lb{reduc}
ds^2_{A}=(\frac{r^2U_{11}}{2})^{1/2}\{ds_{M^4}^2+\frac{dr^2}{h(r)}+\frac{r^2}{2U_{11}}
[detU d\psi^2
+2(U_{11}Q_{\psi}-U_{12}Q_{\phi})d\psi-Q^2_{\phi}+U_{11}H ]\}.
\ee
The components of (\ref{reduc}) are
$$
g_{\psi\varphi}=(\frac{r^2U_{11}}{2})^{1/2}\frac{r^2}{4 U_{11}F\sqrt{\rho}}h(r)[U_{11}sin^2(\theta)
-U_{12}(\frac{\rho}{2}sin(2\theta)sin(\varphi)+\eta sin^2(\theta))]
$$
$$
g_{\psi\theta}=(\frac{r^2U_{11}}{2})^{1/2}\frac{r^2}{4 U_{11}F}h(r)U_{12}\sqrt{\rho}cos(\varphi)
$$
$$
g_{\theta\theta}=(\frac{r^2U_{11}}{2})^{1/2}[\frac{r^2 h(r)}{4}-\frac{r^2 h^2(r)}{8 U_{11}\rho F^2}\rho^2 cos^2(\varphi)]
$$
$$
g_{\varphi\varphi}=(\frac{r^2U_{11}}{2})^{1/2}[\frac{r^2 h(r)}{4}sin^2(\theta)-\frac{r^2 h^2(r)}{8 U_{11}\rho F^2}
sin^2(\theta)(\eta sin(\theta)+\rho cos(\theta)sin(\varphi))^2]
$$
$$
g_{\varphi\theta}=(\frac{r^2U_{11}}{2})^{1/2} \frac{r^2 h^2(r)}{4 U_{11} F^2}
sin(\theta)cos(\varphi)(\eta sin(\theta)+\rho cos(\theta)sin(\varphi)),
$$

where it have been introduced the spherical coordinates $\theta$, $\varphi$ through the relations
$$
u_1=sin(\theta)cos(\varphi),\;\;\;u_2=sin(\theta)sin(\varphi),\;\;\;u_3=cos(\theta).
$$

The range of this coordinates is $\theta \in [0,\pi]$ and $\varphi \in [0,2\pi]$; the other components
of the metric are identically zero.

       The base metric (\ref{quaterna}) have a singularity at $\rho\rightarrow 0$. For this case
it is obtained
$$
U_{11}=\frac{\rho^2(1+3\rho)+\eta^2(9+7\rho)}{2\rho^5} +
 2h(r)[\frac{u_1^2(\rho^2+\eta^2)+(u_2 \eta+u_3 \rho)^2}{\rho^4}]\sim \frac{f(x^i)}{\rho^5},
$$
$$
U_{22}=\frac{8}{\rho^4}+2h(r)(\frac{u_1^2+u_2^2}{\rho^4})\sim \frac{f(x^i)}{\rho^4},
$$
$$
U_{12}=U_{21}=\frac{8\eta}{\rho^4}+h(r)[\frac{(u_1^2+u_2^2)\eta+u_2 u_3 \rho}{\rho^4}]
\sim \frac{f(x^i)}{\rho^4} ,
$$
$$
Q_{\phi}=h(r)\frac{2}{\rho^2}[u_1(\eta du_2+ \rho du_3)-(u_2 \eta+ u_3 \rho)du_1
$$
$$
-u_1(u_3 \eta-u_2 \rho)\frac{2 d\eta}{\rho}]\sim \frac{f(x^i,dx^i)}{\rho^3},
$$
$$
Q_{\psi}=h(r)\frac{1}{\rho^2}(u_1 du_2-u_2 du_1-u_1 u_3 \frac{2 d\eta}{\rho})\sim \frac{f(x^i,dx^i)}{\rho^3},
$$
$$
H=h(r)[|d\overrightarrow{u}|^2+4(u_1^2+u_2^2)\frac{d\eta^2}{\rho^2}
-4\frac{d\eta}{\rho}(u_3 du_2-u_2 du_3)]\sim \frac{f(x^i,dx^i)}{\rho^2}.
$$
where $x^i$ denotes all the coordinates except $\rho$ and the behaviour
for short distances was evaluated. The dilaton field is given explicitly as
$$
\varphi_A=\frac{3}{4}Log\{\frac{r^2}{2}[\frac{\rho^2(1+3\rho)+\eta^2(9+7\rho)}{2\rho^5}]+
r^2h(r)[\frac{u_1^2(\rho^2+\eta^2)+(u_2 \eta+u_3 \rho)^2}{\rho^4}]\}\sim Log(\frac{f(x^i)}{\rho^5}).
$$
The expression for the RR one form is
$$
C=\frac{2\rho\{8\eta+h(r)[(u_1^2+u_2^2)\eta+u_2 u_3 \rho
]\}d\psi}{\rho^2(1+3\rho)+\eta^2(9+7\rho) +
 4\rho h(r)[u_1^2(\rho^2+\eta^2)+(u_2 \eta+u_3 \rho)^2]}
$$
$$
+\frac{4\rho^2 h(r)[\rho u_1(\eta du_2+ \rho du_3)-\rho(u_2 \eta+ u_3 \rho)du_1
-2u_1(u_3 \eta-u_2 \rho)d\eta]}{\rho^2(1+3\rho)+\eta^2(9+7\rho) +
 4\rho h(r)[u_1^2(\rho^2+\eta^2)+(u_2 \eta+u_3 \rho)^2]}\sim f(x^i)\rho.
$$
The components of (\ref{reduc}) diverges in this case for short $\rho$,
$$
g_{\psi\varphi}\sim \frac{f(x^i)}{\rho^5}
,\;\;\;
g_{\psi\theta}\sim \frac{f(x^i)}{\rho^4},\;\;\;
g_{\theta\theta}\sim \frac{f(x^i)}{\rho^2}
$$
$$
g_{\varphi\varphi}\sim \frac{f(x^i)}{\rho^2},\;\;\;
g_{\varphi\theta}\sim \frac{f(x^i)}{\rho}.
$$

   The quaternionic space (\ref{quaternb}) is singular too in the limit $\rho \rightarrow 0$. Using it
as a base space gives
$$
U_{11}=\frac{8\eta^4\rho^2(19+5\rho)+16\eta^6(9+7\rho)+\rho^6(1+35\rho)
+3\eta^2\rho^4(35+61\rho)}{9\rho^5(\rho^2-4\eta^2)^2(8\eta^4+6\eta^2\rho^2+3\rho^4)}
$$
$$
+8h(r)[\frac{u_1^2(\rho^2+\eta^2)+(u_2 \eta+u_3 \rho)^2}{9\rho^4(\rho^2-4\eta^2)^2}]\sim \frac{f(x^i)}{\rho^5} ,
$$
$$
U_{22}=\frac{64(4\eta^4+\rho^4)}{9\rho^4(\rho^2-4\eta^2)^2(8\eta^4+6\eta^2\rho^2+3\rho^4)}
+8h(r)[\frac{u_1^2+u_2^2}{9\rho^4(\rho^2-4\eta^2)^2}]\sim \frac{f(x^i)}{\rho^4},
$$
$$
U_{12}=U_{21}=\frac{32(8\eta^5-4\eta^3\rho^2+3\eta\rho^4)}{9\rho^4(\rho^2-4\eta^2)^2(8\eta^4+6\eta^2\rho^2+3\rho^4)}
+8h(r)[\frac{(u_1^2+u_2^2)\eta+u_2 u_3 \rho}{9\rho^4(\rho^2-4\eta^2)^2}]\sim \frac{f(x^i)}{\rho^4},
$$
$$
Q_{\phi}=4h(r)\frac{1}{3\rho^2(4\eta^2-\rho^2)}\{u_1(\eta du_2+ \rho du_3)-(u_2 \eta+ u_3 \rho)du_1
$$
$$
-u_1(u_3 \eta-u_2 \rho)[\frac{8\eta}{\rho^2-4\eta^2}d\rho+\frac{4(\rho^2-2\eta^2)}{\rho(\rho^2-4\eta^2)}d\eta]\}
\sim \frac{f(x^i,dx^i)}{\rho^3},
$$
$$
Q_{\psi}=h(r)\frac{4}{3\rho^2(4\eta^2-\rho^2)}\{u_1 du_2-u_2 du_1-u_1 u_3 [\frac{8\eta}{\rho^2-4\eta^2}d\rho+
\frac{4(\rho^2-2\eta^2)}{\rho(\rho^2-4\eta^2)}d\eta]\}\sim \frac{f(x^i, dx^i)}{\rho^3},
$$
$$
H=h(r)\{|d\overrightarrow{u}|^2+(u_1^2+u_2^2)[\frac{8\eta}{\rho^2-4\eta^2}d\rho+\frac{4(\rho^2-2\eta^2)}{\rho(\rho^2-4\eta^2)}d\eta]^2
-2[\frac{8\eta}{\rho^2-4\eta^2}d\rho
$$
$$
+\frac{4(\rho^2-2\eta^2)}{\rho(\rho^2-4\eta^2)}d\eta](u_3 du_2-u_2 du_3)\}\sim \frac{f(x^i,dx^i)}{\rho}.
$$
The dilaton field is expressed through the relation
$$
e^{\frac{4}{3}\varphi_D}=\frac{r^2}{2}\{\frac{8\eta^4\rho^2(19+5\rho)+16\eta^6(9+7\rho)+\rho^6(1+35\rho)
+3\eta^2\rho^4(35+61\rho)}{9\rho^5(\rho^2-4\eta^2)^2(8\eta^4+6\eta^2\rho^2+3\rho^4)}
$$
$$
+8h(r)[\frac{u_1^2(\rho^2+\eta^2)+(u_2 \eta+u_3 \rho)^2}{9\rho^4(\rho^2-4\eta^2)^2}]\},
$$
from where follows that
$$
\varphi_D\sim Log[\frac{f(x^i)}{\rho^5}],\;\;\;\; \rho\rightarrow 0.
$$
The behaviour of the RR one-form at short distances results
$$
C \sim f(x^i,dx^i)\rho.
$$
and the components of the IIA metric diverges as
$$
g_{\psi\varphi}\sim \frac{f(x^i)}{\rho^5}
,\;\;\;
g_{\psi\theta}\sim \frac{f(x^i)}{\rho^4},\;\;\;
g_{\theta\theta}\sim \frac{f(x^i)}{\rho^2}
$$
$$
g_{\varphi\varphi}\sim \frac{f(x^i)}{\rho^2},\;\;\;
g_{\varphi\theta}\sim \frac{f(x^i)}{\rho}.
$$

 A detailed analysis of the singularities of the backgrounds corresponding to the $m$-pole solutions and
their physical interpretation was given in \cite{Angelova} and \cite{Angelito}, the interested
reader may consult those references.

   I thanks to A. Isaev for introduce me in the subject, to M.Tsulaia and A.Pashnev for much valuable discussions
and to E.Ivanov for point me out certain features about quaternionic manifolds. Finally, I would like to
thanks L.Masperi, B.Dimitrov and D.Mladenov for constructive critics and encouragement.

\end{document}